\documentclass[a4paper,fleqn,usenatbib]{mnras}

\usepackage{newtxtext,newtxmath}

\usepackage[T1]{fontenc}
\usepackage{ae,aecompl}

\usepackage{graphicx}	
\usepackage{amsmath}	
\usepackage{amssymb}	
\usepackage{enumerate}

\title[Studying the late evolution of a radio-loud AGN in a galaxy group with LOFAR]{Studying the late evolution of a radio-loud AGN in a galaxy group with LOFAR}

\author[F. Savini et al.] {F. Savini$^{1}$\thanks{E-mail: federica.savini@hs.uni-hamburg.de} ,
A. Bonafede$^{1,2}$, 
M. Br{\"u}ggen$^{1}$, 
A. Wilber$^{1}$, 
J. J. Harwood$^{3}$, 
\newauthor 
M. Murgia$^{4}$,
T. Shimwell$^{3,5}$, 
D. Rafferty$^{1}$, 
A. Shulevski$^{3}$,
M. Brienza$^{3,6}$, 
\newauthor  
M. J. Hardcastle$^{7}$,
R. Morganti$^{3,6}$, 
H. R{\"o}ttgering$^{5}$,  
A. O. Clarke$^8$, 
F. de Gasperin$^{5}$, 
\newauthor 
R. van Weeren$^{9}$,
P. N. Best$^{5}$,
A. Botteon$^{2,10}$, 
G. Brunetti$^{2}$, 
R. Cassano$^{2}$
\\
$^1$ Hamburger Sternwarte, Universit\"at Hamburg, Gojenbergsweg 112, 21029, Hamburg, Germany. \\
$^2$ INAF IRA, via Gobetti 101, 40129 Bologna, Italy.\\
$^3$ ASTRON, the Netherlands Institute for Radio Astronomy, Postbus 2, 7990 AA, Dwingeloo, The Netherlands.\\\
$^4$ INAF/Osservatorio Astronomico di Cagliari, Loc. Poggio dei Pini, Strada 54, 09012 Capoterra (CA), Italy.\\
$^5$  Leiden University, Rapenburg 70, 2311 EZ Leiden, Netherlands.\\
$^6$ Kapteyn Astronomical Institute, University of Groningen, PO Box 800, 9700 AV, Groningen, The Netherlands.\\
$^7$ Centre for Astrophysics Research, School of Physics, Astronomy and Mathematics, University of Hertfordshire,\\ 
College Lane, Hatfield AL10 9AB, UK.\\
$^8$ University of Manchester, Jodrell Bank Centre for Astrophysics, Manchester, M139PL, UK. \\
$^9$ Harvard-Smithsonian Center for Astrophysics, 60 Garden Street, Cambridge, 02138, MA, USA.\\
$^{10}$Dipartimento di Fisica e Astronomia, Università di Bologna, via P. Gobetti 93/2, 40129 Bologna, Italy.\\
}

\date{Accepted 2017 November 2. Received 2017 November 2; in original form 2017 August 8}

\pubyear{2017}

\begin{document}
\label{firstpage}
\pagerange{\pageref{firstpage}--\pageref{lastpage}}
\maketitle

\begin{abstract}

Feedback by radio-loud active galactic nuclei (AGN) in galaxy groups is not fully understood. Open questions include the duty cycle of the AGN, the spatial extent of the radio lobes, the effect they have on the intragroup medium, and the fate of the cosmic rays. We present the discovery of a 650 kpc-radio galaxy embedded in steep diffuse emission at $z = 0.18793 \pm 5 \times 10^{-5}$ located at the center of the galaxy group MaxBCG J199.31832+51.72503 using an observation from the LOFAR Two-meter Sky Survey (LoTSS) at the central frequency of 144 MHz. Subsequently, we performed a GMRT observation at the central frequency of 607 MHz to study the spectral properties of the source.  The observations reveal a radio galaxy with a total radio power $P_{\rm tot, 1.4} \sim 2.1 \times 10^{24}$ W Hz$^{-1}$, exhibiting two asymmetrical jets and lobes. The derived spectral index map shows a steepening toward the inner regions and a steep-spectrum core region. We model the integrated radio spectrum, providing two possible interpretations: the radio source is evolved but still active or it is just at the end of its active phase. 
Finally, in the same field of view we have discovered Mpc-sized emission surrounding a close pair of AGN located at a redshift $z = 0.0587 \pm 2 \times 10^{-4}$ (SDSS J131544.56+521213.2 and SDSS J131543.99+521055.7) which could be a radio remnant source.

\end{abstract}

\begin{keywords}
Galaxy clusters; radio emission; individual galaxy cluster:  MaxBCG J199.
\end{keywords}



\section{Introduction}

The majority of galaxies in the local Universe are grouped in dynamically bound systems, such as galaxy groups and clusters. Groups differ from clusters in terms of scaling relations, luminosity functions and halo masses, which are in the range $10^{12}$ - $10^{14} M_{\odot}$ for groups  and $\sim 10^{15} M_{\odot}$ for clusters (e.g. \citealp{Kra2012}). Low-power radio galaxies are commonly found in the centers of rich galaxy groups and clusters, and contribute to the heating of the intra-cluster/-group medium (ICM/IGM) through the on-going activity in their nuclei (\citealp{Croston2005}; \citealp{Croston2014}). Radio-loud active galactic nuclei (AGN) play a crucial role in the thermal evolution of galaxy clusters, providing energy that can offset the radiative losses suffered by the medium and hence averting catastrophic cooling ( \citealp{Fab1991}, \citealp{Sulli2011}). 

The details of this feedback mechanism are still not fully understood, particularly in small galaxy groups, since X-ray measurements are inherently limited to high-temperature groups. Among the many uncertainties about the features of radio galaxies in groups, it is not clear how the AGN affects the thermal state of the intragroup gas and hence the accretion in the nucleus, and how the cosmic rays in the lobes of the radio-loud AGN diffuse, mix, and cool in the IGM (e.g. \citealp{Giaci2011}). The AGN emission can reach hundreds of kpc and in the case of galaxy groups, the low-density environment allows the lobes to expand to scales of up to Mpc (e.g. \citealp{Kaiser1997}, \citealp{Clarke2017}), which has implications for the hydrodynamics of a possible feedback process. Furthermore, the electrons injected into the intragroup medium provide a seed population that could be re-accelerated by shocks and turbulence during group mergers, as occurs in cluster mergers ( e.g. \citealp{vanWeeren2017}).\\

In addition, little is known about what sets the duty cycle of the AGN, i.e. the activity/quiescence phases that radio-loud galaxies undergo. Once the jets of a radio galaxy stop supplying fresh cosmic ray electrons (CRe) to the lobes, the radio sources start to fade on a timescale of $\sim 10^7$ Myr due to losses through synchrotron radiation, inverse Compton, and plasma adiabatic expansion (e.g. \citealp{Kardashev1962}; \citealp{Murgia1999}).  Due to particle energy losses, the high-frequency spectrum steepens with spectral indices\footnote{The spectrum is defined by $S( \nu) \propto \nu^{\alpha}$.} of $\alpha < -1$, and a spectral break develops depending on the magnitude of the energy loss and the age of the particle population.

Low-frequency observations are ideal for discovering steep-spectrum diffuse radio emission since they trace the low-energy, old CRe less affected by the energy losses. \\
The Low Frequency ARray (LOFAR; \citealp{Haarlem2013}) probes the right frequency range and offers the good sensitivity to diffuse, low-surface brightness emission regions due to its {\it uv} plane sampling properties. With its high imaging angular resolution, LOFAR can identify the presence of evolved radio galaxies, explore their morphology, and model their spectrum. With LOFAR, so far, only few remnants have been detected and studied in detail (\citealp{Hard2016}; \citealp{Brie2016}; \citealp{Shulevski2017}). However, the first systematic searches of these sources in the LOFAR fields have already provided an indication of their fraction (between 10 and 30 \%) relative to the entire radio source population (\citealp{BriePro2016}, \citealp{Hard2016}).  
In this paper, we study the peculiar radio source discovered at the center of the galaxy group MaxBCG J199.31832+51.72503 (hereinafter MaxBCG J199) using multi-frequency radio observations to constrain the properties and the origin of the emission discovered by LOFAR. The target was selected from a pointing within the LOFAR Two-meter Sky Survey (LoTSS; \citealp{Shim2017}) after a preliminary inspection of the dataset with the aim of finding radio diffuse emission associated with galaxy groups/clusters. LoTSS is a deep imaging survey carried out as part of the LOFAR Surveys Key Science Project \citep{Rottgering2006} to obtain deep ($\sim$ 100 $\mu$Jy/beam) high-resolution ($\sim$ 5$''$) images at 120 - 168 MHz using the Dutch part of the array in order to map the entire Northern sky.\\

The structure of the paper is the following: we present the source in Sec. \ref{sec:cluster}; we outline the radio observations and data reduction in Sec. \ref{sec:radio}; our main results are presented in Sec. \ref{sec:results}, and we discuss our findings and conclusions in Sec. \ref{sec:disc}, \ref{sec:susp}, and \ref{sec:conc}. 
Throughout the paper, we assume a flat, $\Lambda$CDM cosmology with matter density $\Omega_M = 0.3$ and Hubble constant $H_0 = 67.8$ km s$^{-1}$ Mpc$^{-1}$ \citep{Planck2016}. The angular to physical scale conversion at z = 0.188 is 3.167 kpc/$''$.

\begin{table*}
 \centering
 \caption{ MaxBCG J199 in the Sloan Digital Sky Survey \citep{Koester2007}. 
 Col. 1, Col. 2, Col. 3, Col. 4: Target position, right ascension and declination, longitude and latitude;  Col. 5: Photometric redshift; Col 6: Luminosity in i band;  Col. 7: Number of red-sequence galaxies in the cluster; Col. 8: Radius within which the density of galaxies is 200 times the mean density of such galaxies; Col. 9: $N_{\rm gal}$ within  $R_{\rm 200}$ from the cluster center. Note that $R_{\rm 200}$ and $N_{\rm gal,200}$ are related through a power law (\citealp{Hansen2005}; \citealp{Rykoff2012}).} 
  \begin{tabular}{c c c c c c c c c}
  \hline
 1: RA     &    2: DEC &  3: $l$ & 4: $b$ & 5: $z$  &  6: $L_i$ & 7: $N_{\rm gal}$ &  8: $R_{\rm 200}$ & 9: $N_{\rm gal,200}$\\ 
\scriptsize{(h:m:s, J2000)}			&	 \scriptsize{($^\circ$:$'$:$''$, J2000)}		 &  \scriptsize{($^\circ$)} &  \scriptsize{($^\circ$)} &     &  \scriptsize{($\times 10^{10} L_{\odot}$)}   & & \scriptsize{(kpc)} & \\
13:17:16.4  &  +51:43:30.0  &  199.318  &  51.725 & $0.18 \pm 0.01$ & 17.658 & 13 & 530 & 10\\
\hline
\end{tabular}
\label{info}
\end{table*}

\subsection{The galaxy group MaxBCG J199}
\label{sec:cluster}

The source MaxBCG J199 was classified as a galaxy cluster by \citet{Koester2007} after being identified in the Sloan Digital Sky Survey (SDSS I/II; \citealp{York2000}). The SDSS photometric data were searched for clusters in the redshift range $0.1 \le z \le 0.3$ containing 10 or more red-sequence\footnote{Galaxies with $-24 < M_r < -16$ where $M_r$ is the magnitude in the r band.} galaxies \citep{Bower1992}. The number of galaxies, $N_{\rm gal}$, gives a first estimate of the cluster richness, which is then used to estimate the cluster size $R_{\rm 200} \propto N_{\rm gal}^{0.6}$ Mpc, where $R_{\rm 200}$ is the radius within which the mean density of red-sequence galaxies is $200 \, \Omega_M$  times the mean galaxy density \citep{Hansen2005}. Considering the scaled richness estimate $N_{\rm gal,200}$, i.e. the number of galaxies within $R_{\rm 200}$ from the cluster center, we can use the scaling relation between $M_{\rm 500}$ and $N_{\rm gal,200}$ in \citet{Rozo2009} to get an estimate of the cluster mass (contained within an overdensity of 500 relative to critical at the group redshift): $M_{\rm 500} = e^{B} (N_{\rm gal,200}/40)^A \times 10^{14} M_{\odot}$, where $A = 1.06 \pm 0.17$ and $B = 0.95 \pm 0.16$.
Based on the SDSS selection, the number of galaxies in MaxBCG J199 within $R_{\rm 200}$ is 10, and $M_{\rm 500} = (0.6 \pm 0.2) \times 10^{14} M_{\odot}$. The richness and the estimated mass for MaxBCG J199 are much lower than the typical values for galaxy clusters, hence we will refer to it as a galaxy group.\\ 
Details of the group are summarized in Table \ref{info}.\\

\begin{table*}
\label{obs}
 \centering
 \caption{Details of the radio observations.}
  \begin{tabular}{c c c}
  \hline
Telescope & LOFAR & GMRT  \\
Observation ID & LC2\_038 & 31\_071 \\
Pointing center (RA,DEC) & 13:12:03.2, +52:07:19.4 & 13:17:16.4, +51:43:30.0 \\
Observation date & 2014 Aug 24 & 2016 Dec 25 \\
Total on-source time & 8 h & 8 h \\
Flux calibrator & 3C295 & 3C147 \& 3C286  \\
Central frequency & 144 MHz & 607 MHz \\
Bandwidth  & 48 MHz & 32 MHz \\
Channels & 64 & 256 \\
Integration time & 1 sec & 8 sec\\
Field of view & 5$^\circ$ & 1$^\circ$ \\
Baselines & Dutch (80 - 40000)$\lambda$ & (150 - 49000)$\lambda$\\
\hline
\end{tabular}
\label{obs-data}
\end{table*}

\section{Radio observations and data reduction}
\label{sec:radio}

Summaries of the observations can be found in Table \ref{obs-data}.\\
The calibration and imaging procedure performed on the Low Frequency Array (LOFAR) and Giant Meter Radio Telescope (GMRT) observations are outlined below. 
We consider a calibration error of $15\%$ on all the measured flux densities (\citealp{Shi2016}, \citealp{vanWeeren2016b}).

\subsection{LOFAR}

LOFAR is an array of antenna dipoles grouped into so-called stations (see \citealp{Haarlem2013} for details). The Low Band Antennas (LBA) operate in the range 10 - 90 MHz and the High Band Antennas (HBA) in the range 110 - 240 MHz.\\
In this paper, we present a HBA LOFAR observation at the central frequency of 144 MHz within LoTSS. A preliminary pre-processing step has been performed through a pipeline offered by the Radio Observatory (ASTRON) to flag bad data and average in time and frequency (down to 0.1 MHz/ch and 8 s). Data reduction was performed following the calibration scheme described in \citet{vanWeeren2016}, which has been developed to correct for direction-dependent effects within the observed field of view at HBA frequencies. The calibration scheme consists of two main components: a non-directional part and a directional part, briefly summarized below. For more details we refer the reader to \citet{vanWeeren2016}.

\subsubsection{Pre-Facet Calibration }
The direction-independent part, so-called Pre-Facet Calibration (Prefactor pipeline\footnote{https://github.com/lofar-astron/prefactor}), is a preparatory step for the directional calibration processing. Amplitudes and phase gains, station phase correlation offsets, and clock-TEC\footnote{TEC refers to the station differential Total Electron Content.} solutions are calculated for the flux calibrator, adopting the flux scale of \citet{SH2012}. The flux calibrator for our dataset is 3C295 and was observed for 10 minutes.\\
After these steps we transferred the amplitude gains, station phase correlation offsets, and clock offset to the target data. 
An initial phase calibration was performed using a low-resolution sky model (Global Sky Model for LOFAR\footnote{https://www.astron.nl/radio-observatory/lofar/lofar-imaging-cookbook}) from the VLA Low-Frequency Sky Survey Redux (VLSSr ; \citealp{Lane2012}), the Westerbork Northern Sky Survey (WENSS; \citealp{Renge1997}), and the NRAO VLA Sky Survey (NVSS; \citealp{Condon1998}).\\ 
High-resolution (39$'' \times$ 31$''$) and low-resolution (126$'' \times$ 108$''$) direction-independent calibrated images were obtained through a step called Initial Subtraction. 

In this step, high-resolution compact sources are masked and imaged. Their clean components are then subtracted from the {\it uv} data and listed into a sky model (one for each subband). Diffuse emission that was not visible in the high-resolution images can now be detected and low-resolution sources are then masked and imaged. The low-resolution components are also subtracted from the {\it uv} data and then added to the sky model. 
The calibrator 3C295, which appears as a bright source far outside the FWHM of the primary beam ($\sim 8^\circ$ far away from the science target) causing some artifacts in the field of view, was peeled off from the first half of the bandwidth where the effects are more relevant.

\begin{figure}
\centering
\includegraphics[width=0.50\textwidth]{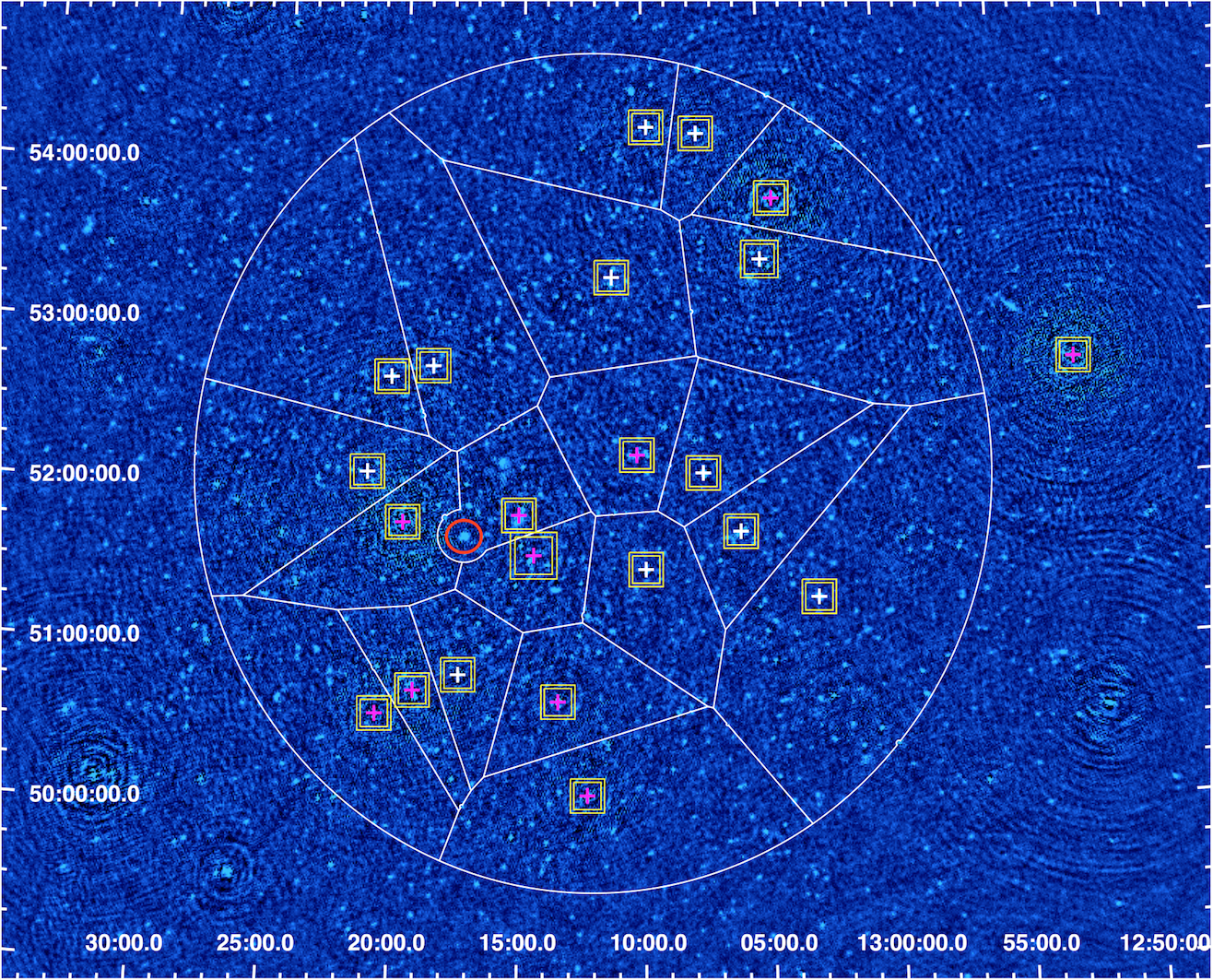}
\caption{Wide-field high-resolution image with the overlay of the facet layout generated by the Factor pipeline. White regions show the facets and the elliptical region that encompasses the faceted area with a $2.5^\circ$ radius adjusted for the primary beam shape. Outside this radius, only small patches, which are faster to process, are used. The coordinates of the target were specified in the parameter settings to include the source in one single facet that therefore shows a curved boundary. Each calibrator is indicated with a cross (magenta for the facets that were processed with Factor and white for the non-processed facets) and the region used in the self-calibration step is indicated with a yellow square. The target is indicated with a red circle.}
\label{facets}
\end{figure}

\begin{table*}
\label{obs}
 \centering
  \caption{
 Col. 1: Telescope/Survey; Col. 2: Central frequency; Col. 3: Minimum baseline; Col. 4: Largest angular scale; Col. 5: Resolution; Col 6: rms noise; Col. 7: Parameters used for LOFAR and GMRT imaging, such as taper (T) and weighting scheme; when Briggs weighting scheme is used, the robust value is specified \citep{Briggs}.}
 \begin{tabular}{c c c c c c c}
  \hline
1: Telescope & 2: Freq. &  3: $B_{min}$ & 4: $LAS$ & 5: Res. & 6: rms & 7: Imaging\\
 & \scriptsize{(MHz)} &  \scriptsize{($\lambda$)} & & & \scriptsize{(mJy/beam)} & \\
 LOFAR & 144 & 80  & 2578$''$ & $10.6'' \times 6.0''$  & 135 & Briggs -0.25\\
  & & 150 & 1375$''$ & $19'' \times 19''$  & 350 & uniform, 15$''$ T\\
 &  & 80 & &$27'' \times 26''$   & 350 & Briggs 0, 20$''$ T\\
  GMRT & 607 & 150  & 1375$''$ & $6.0'' \times 4.8''$ & 60 & Briggs -0.25\\
   &  & 150  & &$18'' \times 17''$ & 250 & Briggs 0, 20$''$ T\\
   &  & 150 &  &$19'' \times 19''$ & 290 & uniform, 20$''$ T\\
 VLSSr & 74 & 94 & 1100$''$ & $80'' \times 80''$  & 50000 & -\\
 WENSS & 325 &  150 & 1375$''$ & $54'' \times 68''$  & 300 & -\\
 NVSS & 1400 & 210 & 970$''$ &$45'' \times 45''$  & 500 & -\\
 FIRST & 1400 & 1720 & 60$''$ &$5.4'' \times 5.4''$  & 200 & -\\

\hline
\end{tabular}
\label{results}
\end{table*}

\subsubsection{Facet Calibration}
The direction-dependent step, so-called Facet-Calibration (Factor pipeline\footnote{https://github.com/lofar-astron/factor}), is based on dividing the sky into a discrete number of directions (facets) covering the observed field of view and calibrating each of these directions separately. The aim is to calculate the direction-dependent corrections needed to obtain near-thermal-noise-limited images using the full resolution offered by LOFAR.  
The input needed for the pipeline are the sky models obtained through Initial Subtraction and empty datasets that will be filled with the calibrated sources. A default calibrator, typically a bright compact source, is selected for each facet with restrictions, such as minimum flux density in the highest-frequency band and maximum size. The user may modify the calibration region, and multiple sources within the region can be used. We use settings in the Factor pipeline to restrict the facet calibrator sources to have a reasonable total number of facets with reasonable sizes. The aim is to compute solutions on small portions of the sky and reduce the processing time. We choose a minimum flux density of 0.6 Jy for the calibrators, and we also choose to calibrate and image using baselines above 80 $\lambda$ to prevent residual diffuse emission seen by the shortest baselines from affecting the results. The coordinates of MaxBCG J199 and a 10$'$ radius around it are also specified to include the source in one single facet. After performing self-calibration cycles on the calibrator, all the fainter sources in the facet are added back and calibrated using the calculated solutions which are assumed to apply to the whole facet. An updated sky model for the region of the sky covering the facet is obtained and then subtracted from the {\it uv} data and the whole process is repeated to finally obtain a direction-dependent corrected image for each facet.
Finally, a mosaic field image containing all the facets' images is corrected for the primary beam. Re-imaging was performed with different parameters on the target facet to obtain different resolutions and weights to increase sensitivity to diffuse, extended emission. 
As we were focusing on one single science target, we chose to process only 9 facets, i.e. the brightest sources in the field and those bordering the target facet. In addition, the last and 10th facet we processed was the target facet itself, so that it could benefit from the improved subtraction obtained by calibrating the preceding facets. The calibration regions of the processed facets are indicated with yellow boxes in Fig. \ref{facets}.

\subsection{GMRT}
We used a follow-up GMRT observation in the range 591 - 623 MHz to enable a study of the spectral properties of the sources.\\ 
In the GMRT calibration the sources 3C147 and 3C286 were used as absolute flux and bandpass calibrators respectively and were observed for 10-15 minutes, at the beginning and at the end of the target observation. The source 1400+621 was used as a phase calibrator and was observed every 10-15 minutes. 
Data reduction was performed using the CASA tools (Common Astronomy Software Applications, version 4.5.2; \citealp{Mc2007}). After inspecting the dataset, bad data were flagged through both manual flagging and using the AOFlagger software \citep{Off2012}. Flux and bandpass calibration were performed against 3C147 and 3C286, adopting the flux scale in \citet{PB2013}\footnote{We used the most updated flux scale, although different from the scale used for LOFAR observations, since the flux difference between the Perley-Butler 2013 and Scaife-Heald models is within the calibration error of 15\%.}. Gain phases and amplitudes were calibrated every 10-15 minutes against 1400+621. One compact source (13:09:46, +51:48:10) residing in the primary lobes of the primary beam was peeled off. To speed up the imaging process, the dataset has been averaged in frequency and time (down to 3.2 MHz/ch and 16 s). Imaging was carried out in CASA, using the multi-frequency synthesis (MFS) CLEAN algorithm \citep{Rau2011} and the wide-field imaging technique to compensate for the non-coplanarity of the array. 

Only one cycle of phase self-calibration cycle was needed to reach convergence and obtain an image of the target field, which was finally corrected for the primary beam.

\begin{figure*}
\centering
\includegraphics[width=1.1\textwidth]{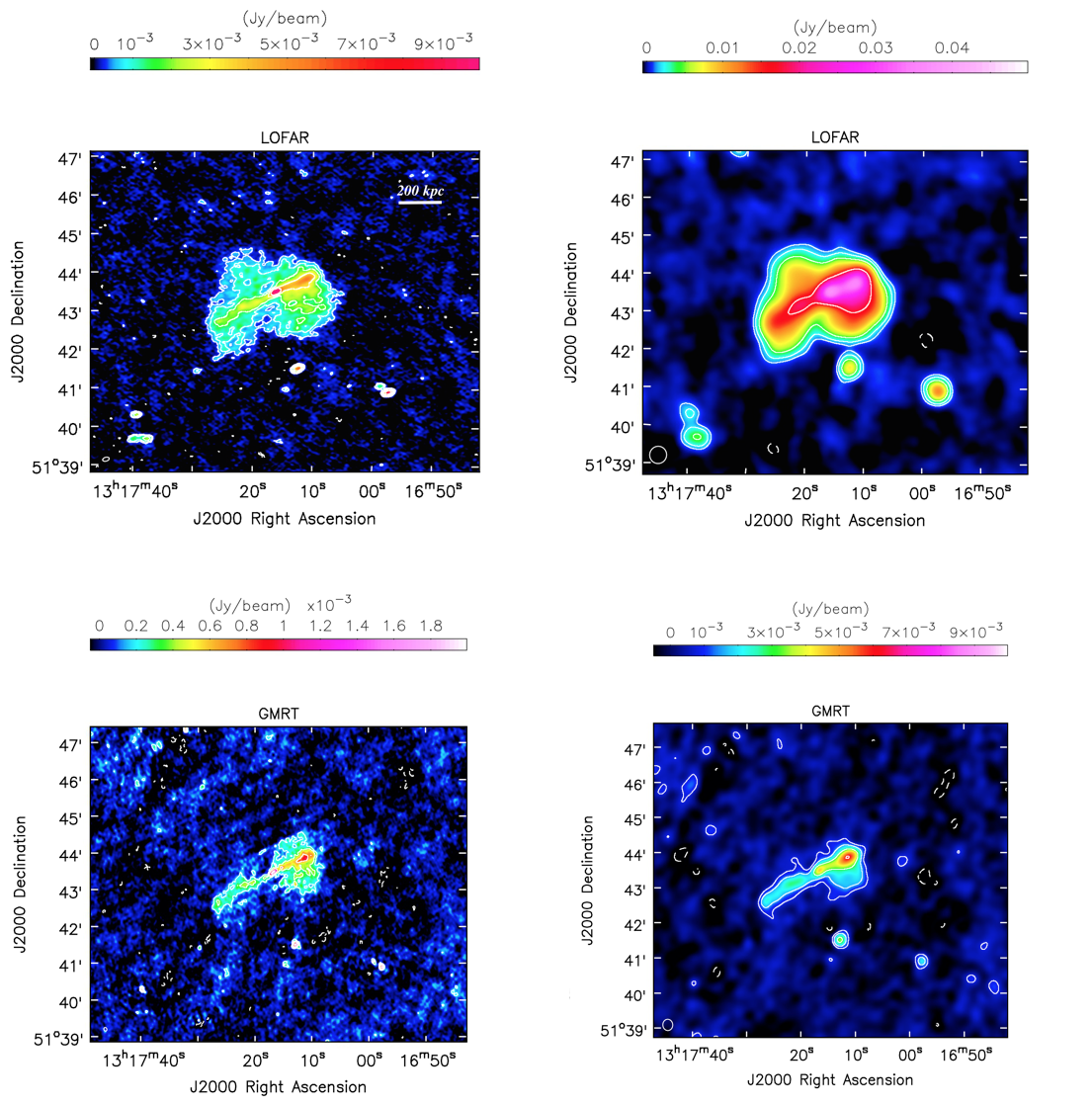}
\caption{ {\bf Top left:} LOFAR image of MaxBCG J199 at  the central frequency 144 MHz obtained with the Briggs scheme \citep{Briggs}, robust=-0.25, and no taper. The contour levels are at $(-1, 1, 2, 4, 8, 16)\, \times \, 3\sigma$ where $\sigma$ = 135 $\mu$Jy/beam. The beam shown at the bottom left of the image is 10.6$'' \times$ 6.0$''$.  The image shows the presence of a radio galaxy embedded in radio diffuse emission with maximum angular size of $3.4'$ which corresponds to projected linear size of 650 kpc. {\bf Top right:} LOFAR image of MaxBCG J199 at  the central frequency 144 MHz obtained with the Briggs scheme \citep{Briggs}, robust=0, and taper of 15$''$. The contour levels are at $(-1, 1, 2, 4, 8, 16)\, \times \, 3\sigma$ where $\sigma$ = 350 $\mu$Jy/beam. The beam shown at the bottom left of the image is 27 $'' \times$ 26$''$. No additional diffuse emission can be observed in this image. {\bf Bottom left:} GMRT image of MaxBCG J199 at  the central frequency 607 MHz obtained with the Briggs scheme \citep{Briggs}, robust=0, and no taper. The contour levels are at $(-1, 1, 2, 4, 8, 16)\, \times \, 3\sigma$ where $\sigma$ = 60 $\mu$Jy/beam. The beam shown at the bottom left of the image is  6.0$'' \times$ 4.8$''$.  The image shows the presence of the jets of the radio galaxy. {\bf Bottom left:} GMRT image of MaxBCG J199 at  the central frequency 607 MHz obtained with the Briggs scheme \citep{Briggs}, robust=-0.25, and taper of 20$''$. The contour levels are at $(-1, 1, 2, 4, 8, 16)\, \times \, 3\sigma$ where $\sigma$ = 250 $\mu$Jy/beam. The beam shown at the bottom left of the image is 19$'' \times$ 17$''$. No diffuse emission can be observed in this image.}
\label{4}
\end{figure*}

\begin{figure*}
\centering
\includegraphics[width=1\textwidth]{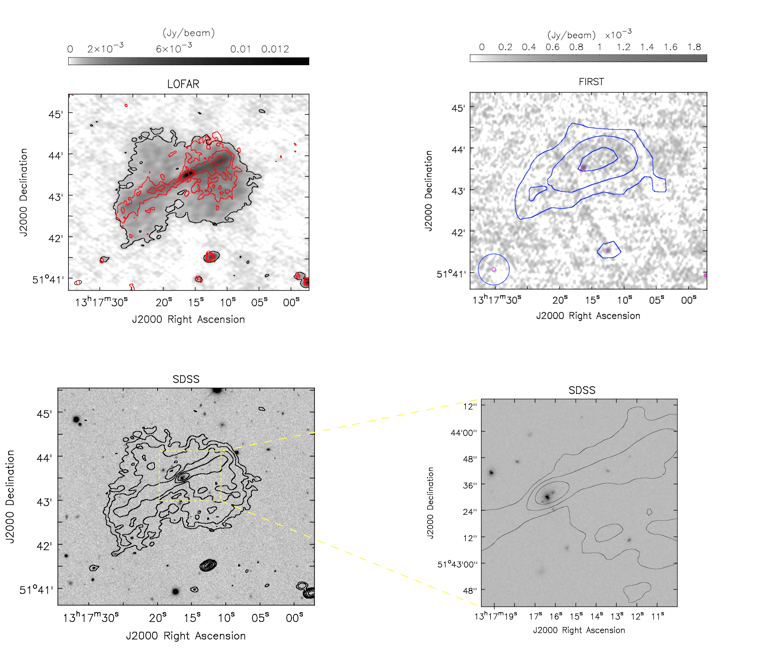}
\caption{ {\bf Top left:} LOFAR image of MaxBCG J199 at  the central frequency  144 MHz in greyscale with its contour level at $3\sigma$ where $\sigma$ = 135 $\mu$Jy/beam in black. GMRT 607 MHz contours levels at $(1, 2, 4, 8, 16)\, \times \, 3\sigma$ where $\sigma$ = 60 $\mu$Jy/beam are overlaid in red. The beams shown at the bottom left of the image are  10.6$'' \times$ 6.0$''$ and 6.0$'' \times$ 4.8$''$ for LOFAR and GMRT respectively.  {\bf Top right:} FIRST image of MaxBCG J199 at 1.4 GHz in greyscale with its contour level at $3\sigma$ where $\sigma$ = 200 $\mu$Jy/beam in magenta. NVSS contours levels at $(1, 2, 4)\, \times \, 3\sigma$ where $\sigma$ = 500 $\mu$Jy/beam are overlaid in blue. The beams shown at the bottom left of the image are $5.4'' \times$ 5.4$''$ and 45$'' \times$ 45$''$ for FIRST and NVSS respectively.  With the FIRST snapshot observation only the core region of the radio galaxy is visible, whereas NVSS observation could detect radio emission but without the resolution required to distinguish internal features. {\bf Bottom left:} SDSSg,r,i mosaic image in greyscale with LOFAR 144 MHz contour levels at $3\sigma$ where $\sigma$ = 135 $\mu$Jy/beam in black and GMRT 607 MHz contour levels at $3\sigma$ where $\sigma$ = 60 $\mu$Jy/beam in red. The radio diffuse emission surroundings the two jets can be seen only with LOFAR. {\bf Bottom right:} zoom of the SDSSg,r,i mosaic image in the core region. LOFAR 144 MHz contour levels at $(4, 8, 16)\, \times \, 3\sigma$  where $\sigma$ = 135 $\mu$Jy/beam in black. Three galaxies can be seen corresponding to the inner region of the radio emission. }
\label{4mix}
\end{figure*}

\section{Results}
\label{sec:results}

To study the emission and its spectral properties, we have made several images at high- and low- resolution, summarized in Tab. \ref{results} and shown in Fig. \ref{4}. The images obtained for spectral analysis (see Sec. \ref{spec}) are not shown.\\ 
The LOFAR images show that the radio emission associated with MaxBCG J199 is coming from a radio galaxy with a pair of jets and lobes extending from a compact core, and diffuse emission likely connected to the AGN. 
The core of the radio galaxy is coincident with the optical source SDSS J131716.39+514330.1, which corresponds to the brightest galaxy of the group and is identified as a broad-line galaxy with a redshift of $z = 0.18793 \pm 5 \times 10^{-5}$ (Data Release 13; \citealp{SDSS2016}). However, the structure of the core region appears to be more complex, since SDSS images show three galaxies which create a multiple core system, as visible in the bottom right panel of Fig. \ref{4mix}. Moreover, there is an offset (smaller than beam in the full-resolution LOFAR image,) between the peak of the radio emission and the central galaxy seen in the optical image.\\
The radio diffuse emission has a mean surface brightness of 4.5 $ \mu$Jy/$''$$^2$ at 144 MHz and an extent of $3.4'$, which corresponds to a projected linear size of 650 kpc at the spectroscopic redshift of the BCG\footnote{The photometric redshift of the galaxy group ($z = 0.18  \pm 0.01$) given in \citet{Koester2007} is not used throughout the paper, instead we use the spectroscopic redshift of the BCG ($z = 0.18793 \pm 5 \times 10^{-5}$) given by \citet{SDSS2016} since the radio emission observed is related to that galaxy.}. Only the brightest and more compact emission is visible in the GMRT images at the achieved sensitivity level.\\
We note an asymmetry in the jet intensity and morphology: the NW lobe is brighter and it extends across the direction of the jet axis with a wing in the south-west direction, whereas the SE lobe is fainter, fading away towards the edge, but with a larger projected linear size. \\

Radio emission at the group coordinates is observed in the following surveys: NVSS at 1.4 GHz, WENSS at 325 MHz, and the VLA survey at 1.4 GHz, Faint Images of the Radio Sky at Twenty Centimeters (FIRST, \citealp{Beck1995}). NVSS and WENSS are sensitive to emission from sources extended on scales of arcminutes, but their resolution ($45'' \times 45''$ and $54'' \times 68''$ at the group declination, respectively) and sensitivity (500 $\mu$Jy/beam and 300 $\mu$Jy/beam, respectively) are too low to identify features, such as a core or jet emission, in the radio emission. With FIRST, which is a 3-min snapshot with resolution of 5$''$, only the region closest to the core is visible. The FIRST image with an overlay of NVSS contours is shown in the top right panel of Fig. \ref{4mix}. In addition, the source is detected in the Green Bank 6 cm survey at 4.85 GHz (GB6; \citealp{Greg1996}), and in the VLSSr at 74 MHz. The flux densities from these survey observations can be combined to obtain an integrated radio spectrum (see Sec. \ref{spec}).\\
No information about the X-ray emission of the group is available in the literature. The group is not detected in the ROSAT All Sky Survey (RASS; \citealp{Voges1999}) and no pointed observations exist.\\

From the NVSS image we measured the integrated flux density at 1.4 GHz to be $S_{\rm tot, 1.4} = 21 \pm 3$ mJy, corresponding to a total radio power of $P_{\rm tot, 1.4} \sim 2.1 \times 10^{24}$ W Hz$^{-1}$. \\ 
The core is resolved at 1.4 GHz by FIRST and we measure the integrated flux density of the core region to be $S_{\rm core, 1.4} = 1.3 \pm 0.2$ mJy.

\begin{figure}
\centering
\includegraphics[width=0.5\textwidth]{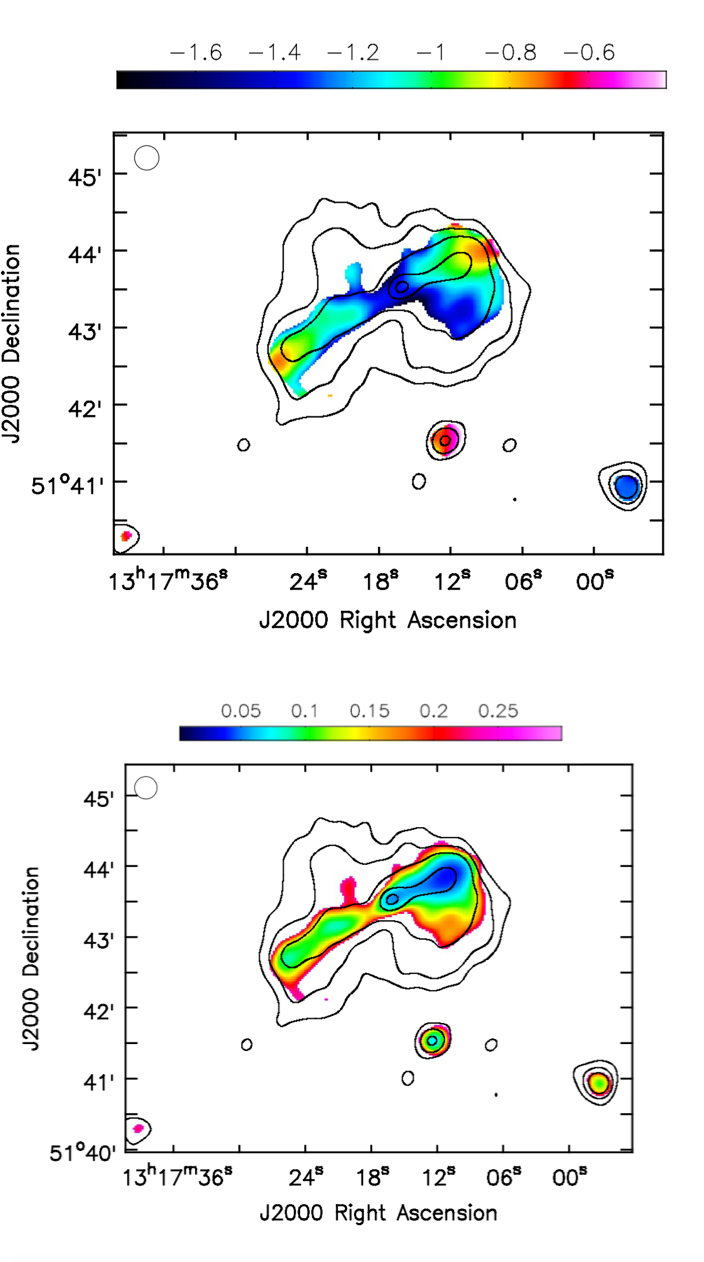}
\caption{Spectral index map (top) and relative error map (bottom) between the 144 MHz LOFAR image and the 607 MHz GMRT image with an overlay of LOFAR contour levels at $(3, 10, 20, 50, 70)\, \times \, \sigma$ where $\sigma$ = 350 $\mu$Jy/beam. The beam shown at the top left of the image is $19'' \times 19''$ .}
  \label{spix}
\end{figure}

\subsection{Spectral analysis}
\label{spec}

In order to study the spectral properties of MaxBCG J199, we have reimaged LOFAR and GMRT data with a resolution of 19$''$, and same pixel size, baseline range (150 - 49000 $\lambda$) and uniform weighting scheme to minimize the effects of differences in the {\it uv} coverage of the two interferometers. We have produced a low-frequency spectral index map using CASA tasks, shown in Fig. \ref{spix}. The spectral index values are calculated in the region where both LOFAR and GMRT images are above 3$\sigma$, where $\sigma$ is 290 $\mu$Jy/beam and 350 $\mu$Jy/beam for GMRT and LOFAR respectively. Pixels below 3$\sigma$ are blanked out. The spectral index error map is obtained using the following equation:

\begin{equation}
\Delta \alpha = \frac{1}{\log \frac{\nu_1}{\nu_2}} \sqrt{ \Big( \frac{\Delta S_1}{S_1} \Big)^2 + \Big( \frac{\Delta S_2}{S_2} \Big)^2} ,
\end{equation}
where $S_1$ and $S_2$ are the flux densities at frequencies $\nu_1$ and $\nu_2$ and $\Delta S_1$ and $\Delta S_2$ are the respective errors which include the measured map noises and flux calibration errors.\\ 
The spectral index values range from -1.3 to -1.1 in the core region and inner edges, and from -0.7 to -0.5 at the outer edges. The lobes have spectral index values that flatten towards the outermost lobe edges; this is especially prominent in the NW lobe. We note that the regions that show the flattest  spectral indices, which are also regions with large errors, do not match the highest surface brightness regions in the LOFAR and GMRT maps. \\

\begin{figure}
\centering
\includegraphics[width=0.5\textwidth]{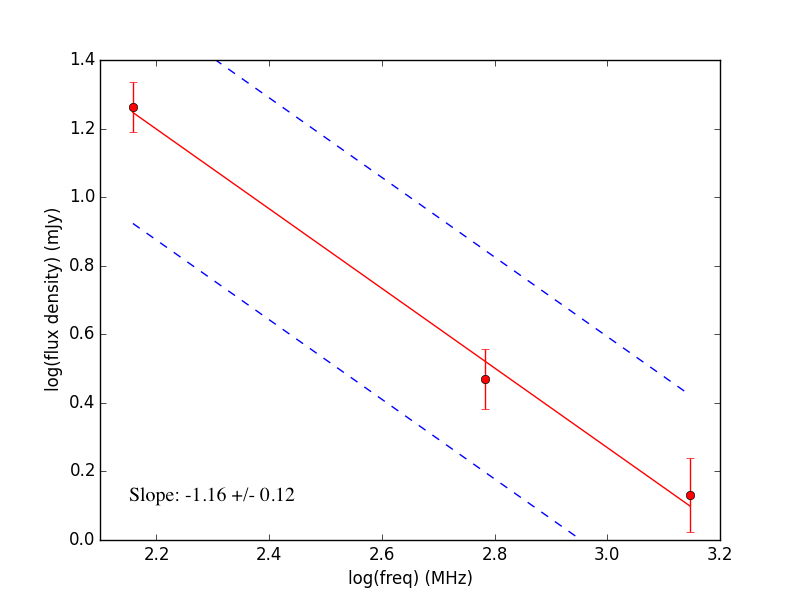}
\caption{Plot of flux density against frequency with linear fit overlaid for LOFAR, GMRT and FIRST measurements of the core region of MaxBCG J199.} 
\label{fit}
\end{figure}

The global spectral index value of the source calculated using the integrated flux densities from the area where the GMRT detection is above 3$\sigma$ is relatively steep: $\alpha_{144}^{607}$ = $-1.14 \pm 0.13$.\\ 
The observed LOFAR emission extends well beyond the GMRT emission and the diffuse emission detected at 144 MHz that can not be seen at 607 MHz must be much steeper. This emission is detected by LOFAR even when only the same baselines as the GMRT are imaged. 
This enables us to derive a spectral index upper limit to the value of $\alpha_{144}^{607} < -1.8 \pm 0.2$, considering the LOFAR integrated flux density of the lobe and a $3\sigma$ GMRT flux density upper limit where $\sigma$ was determined from a set of flux density measurements in the radio galaxy region.\\

The spectral index map reveals that the emission closest to the radio galaxy core is steep with $\alpha_{\rm 144}^{\rm 607} < -1$. The core can be seen in the intensity images also at 1.4 GHz (FIRST). Therefore, we can obtain a three-point spectrum of the core region, using LOFAR, GMRT, and FIRST images readjusted to have the same {\it uv} range, and same beam (6$''$) \footnote{We used a uniform weighting scheme to image GMRT and LOFAR datasets, whereas we note that the FIRST image could have been obtained with a different scheme. However, the effects of different weighting schemes should not be relevant since only the core region is visible in the FIRST image and it is unresolved.}. We measured the flux density from a region corresponding to the brightest central emission in each map and respective errors that include the measured map noises and flux calibration errors and estimated the spectral index of the core to be $\alpha_{\rm 144,607,1400}= -1.16 \pm 0.12$ performing a linear fit, as shown in Fig. \ref{fit}. \\

\begin{table}
\label{obs}
 \centering
  \caption{Flux densities from VLSSr, LOFAR, WENSS, GMRT, NVSS, and GB6 in the region of MaxBCG J199. All the measurements onto the absolute flux density of \citet{Ba1977}.}
  \begin{tabular}{c c c}
  \hline
Central frequency & Flux density & Error\\
\scriptsize{(MHz)} & \scriptsize{(mJy)} & \scriptsize{(mJy)} \\
 
74   &  	823.0	&	214.0\\
144 &   	388.0	&	38.8\\
325&	115.5 	&	14.0	\\	
607	&       93.7	&	9.4\\
1400   &  25.7	&	2.6\\
4850    & 2.9	&	0.3\\
\hline
\end{tabular}
\label{fluxes}
\end{table}

\begin{figure}
        \centering
 \includegraphics[width=8cm]{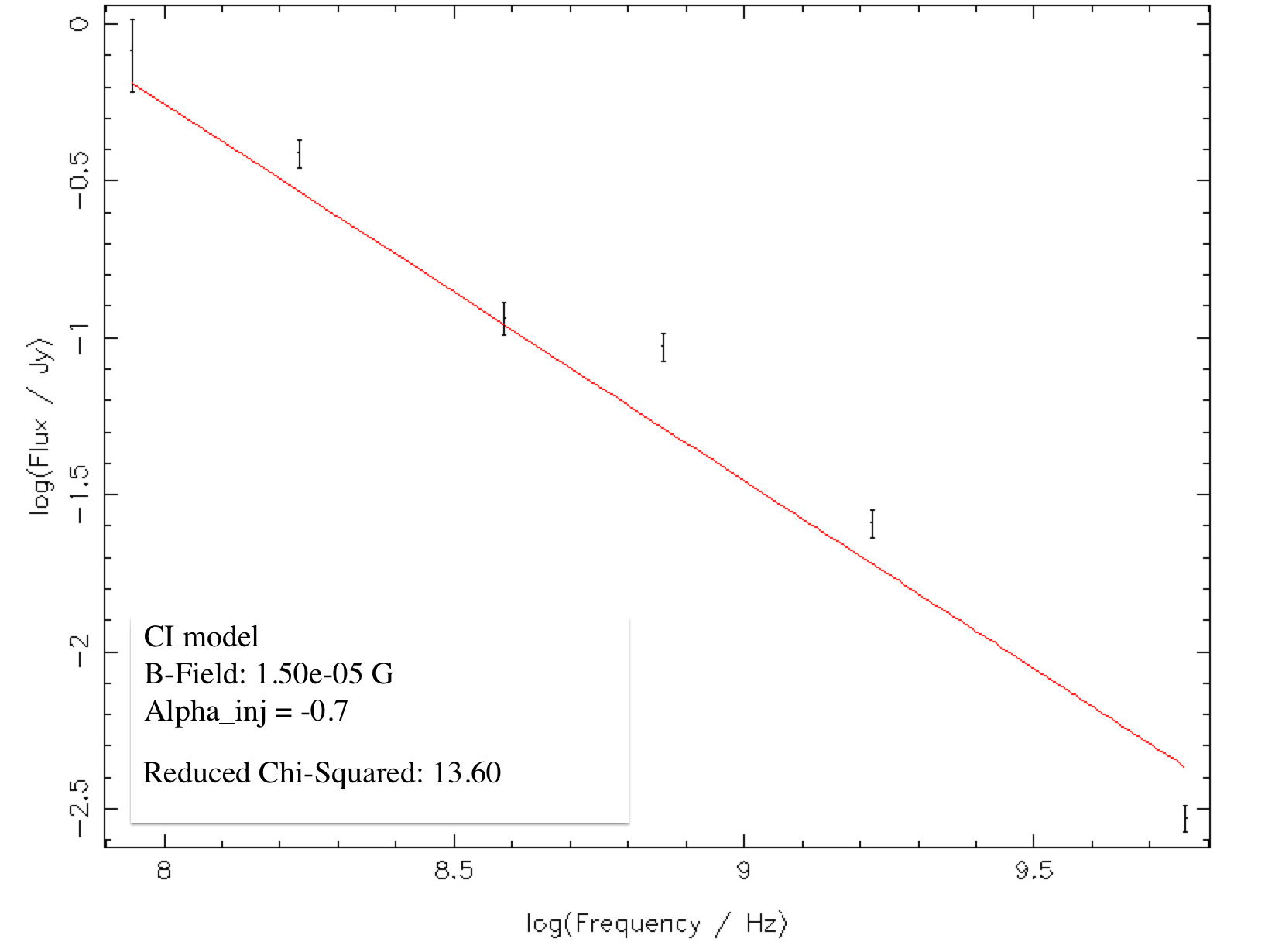} \quad\includegraphics[width=8cm]{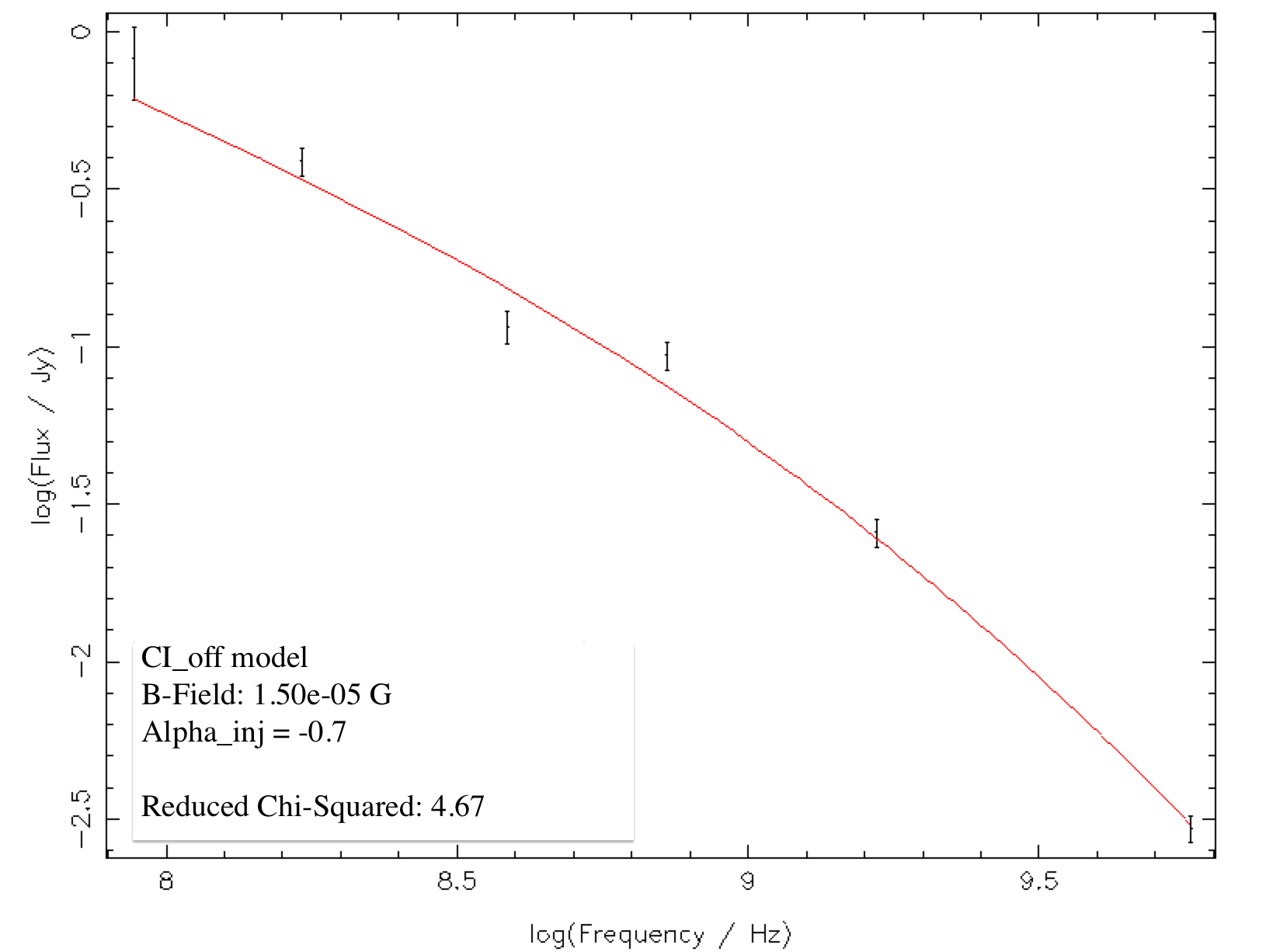} 

\caption{Fit and best-fit parameters of the CI model and CI$_{\rm off}$ model of the integrated radio spectrum.}          
 \label{integ}
\end{figure}

In addition to the spectral information derived from our observations with LOFAR and GMRT, we measured the flux densities from {\bf VLSSr}, WENSS, NVSS, and GB6 in the region of MaxBCG J199 constrained by the $3\sigma$ LOFAR contours to obtain the integrated radio spectrum of the source shown in Fig. \ref{integ}. In Tab. \ref{fluxes} we report all the measurements. They were placed onto the absolute flux density of \citet{Ba1977} by scaling for the multiplicative factor listed in \citet{He2008}.\\ 

We computed the best-fit synchrotron model of the spectrum using the Broadband Radio Astronomy ToolS (BRATS\footnote{http://www.askanastronomer.co.uk/brats};  \citealp{Har2013}, \citealp{Har2015}) software package, comparing two models:
\begin{itemize}

\item the continuous injection (CI; \citealp{JP1973}) model for active sources, which assumes that fresh electrons are injected at a constant rate for a duration $t_{\rm CI}$ \citep{Pa1970}. When the source is active, its radio spectrum changes as a function of time $t$, and the break frequency $\nu_b$ shifts to lower values, via 

\begin{equation}
\nu_b \propto \frac{B} {t^2 \, (B^2 + B_{\rm IC}^2)} ,
\end{equation}

where $B$ is the magnetic field, and $B_{\rm IC}$ is the equivalent magnetic field due to inverse Compton scattering of cosmic microwave background photons.\\

\item the CI$_{\rm off}$ model (\citealp{KG1994}; \citealp{Murgia2011}), which extends the Jaffe \& Perola model \citep{JP1973} to inactive sources. When the electron supply stops, the source enters the quiescence phase of duration $t_{\rm off}$ and the synchrotron age is $t_s = t_{\rm CI} + t_{\rm off}$.  The break frequency $\nu_{\rm b, off}$ evolves via

\begin{equation}
\nu_{\rm b, off} = \frac{\nu_b \, (t_{\rm off} + t_{\rm CI})^2} {t_{\rm off}^2} .\\ \\
\end{equation}
\end{itemize}

To estimate the magnetic field, we made the simple assumption of equipartition between relativistic particles and a uniform magnetic field and we calculated the minimum energy density $u_{\rm min}$ and the equipartition magnetic field $B_{\rm eq}$ for MaxBCG J199 using the revised formula in \cite{BK2005}. We adopted the source flux density at 144 MHz, where the energy losses of the synchrotron electrons ($\propto E^2$) should be negligible, the global spectral index value computed in the previous section ($\alpha \sim -1.1$), an electron/proton ratio of 100, and a volume filling factor of 1. Moreover, we assumed ellipsoidal geometry, hence a value of 200 kpc for the source depth. The resulting value is $B_{\rm eq}$[$\mu$G] $\sim$ 15.\\

The fit to the models and the best-fit parameters are shown in Fig. \ref{integ}. \\
For both models, we assumed that the injected particles have a power-law energy spectrum $N(E) \propto E^{\delta}$ which results in a power-law radiation spectrum with spectral index $\alpha_{\rm inj} = (\delta + 1)/2$ over a wide range of frequencies. For the injection spectral index, we assumed $\alpha_{\rm inj}$ = -0.7, which is the value measured in the flattest regions of the source. Fixing $\alpha_{\rm inj}$ helps us to limit the number of free parameters for the model. We also neglect adiabatic losses and assume that the pitch angles of the radiating electrons are continually isotropized in a time much shorter than the radiative timescale, which implies that the synchrotron energy losses are the same for all electrons. \\

Using an estimate of the magnetic field strength $B$[$\mu$G] and the break frequency $\nu_b$[GHz] obtained from the spectral fitting of CI and CI$_{\rm off}$ models, it is possible to derive the spectral age of the source \citep{Murgia2011}, via

\begin{equation}
t_{\rm s} [{\rm Myr}]= 1590 \, \frac{B^{0.5}}{( B^2 + B^2_{\rm IC} ) \, [ (1 + z)\, \nu_b ]^{0.5}} . \\ \\
\end{equation}

Assuming the source magnetic field $B$ constant and equal to the value computed using the above equipartition approach, the characteristic spectral age can be calculated for both models.\\
The best fit for the CI model is found for the break frequency value $0.6$ MHz. However, the age that can be derived is very poorly fitted ($t_{\rm CI} \sim 830$ Myr), since the fit is forced to be as steep as possible before being limited by the low-energy cut off. \\
The best fit for the CI$_{\rm off}$ model is found for the break frequency value $439$ MHz with an off component break at $12$ GHz. The time during which the source has been on and off are estimated to $t_{\rm CI} \sim 25$ and $t_{\rm off} \sim 6$ Myr respectively, which give a total age of $t_{\rm s} \sim 31$ Myr.\\
As recently demonstrated by \citet{Har2017}, though, the CI and CI$_{\rm off}$ models are unable to provide a robust measure of the source's spectral age. Therefore, the break frequencies derived by modeling the integrated spectrum are to be considered only an indication of the break frequencies of the source, whereas they can provide a potentially useful tool for discerning between active and remnant radio galaxies. Possible interpretations of the origin of this source will be investigated in Sec. \ref{sec:disc}.

\section{Discussion}
\label{sec:disc}

At LOFAR frequencies it was possible to resolve the inner structure of the radio emission of the galaxy group MaxBCG J199, hosting a central radio galaxy which shows opposing radio jets, and lobes extending from the core out to hundreds of kpc. The spectrum of the brightest galaxy of the group indicates a broad-line galaxy (Data Release 13; \citealp{SDSS2016}), therefore we expect an inclination of the AGN with respect to the line of sight, i.e. the jets are not in the plane of the sky. Moreover, the structure of the jets clearly shows an asymmetry in brightness that is a further indication that the source is inclined.
The jet pointing towards NW is likely to be directed towards the observer, since the intensity of the approaching jet is enhanced as a consequence of bulk relativistic motion (Doppler boosting effect; \citealp{RL1979}). Analyzing the statistical distribution of the broad-line AGN orientations, \citet{Marin2016} placed the inclination angle between the jets and the line of sight in the range $0^{\circ} - 70^{\circ}$, with a mean expected value of $33^{\circ}$.\\

Luminous jets are typical of FR-I sources since the energy transport from the core to the edges is inefficient due to radiative losses and interaction with the surrounding environment. On the other hand, the spectral index map for this source suggests that there might be electron acceleration at the lobe edges in regions known as hotspots, and a consequential backflow typical of FR-II sources: the two jets are bent by interaction with the intragroup medium and the plasma is aging while traveling back toward the core region. However, clear hotspots are not visible in the intensity maps. As the classification scheme is purely morphological, other observed features, such as the power of the radio galaxy, cannot provide a definitive test of the source's FR type. The asymmetry in the jets' brightness suggests Doppler boosting that requires relativistic electrons, and usually the most powerful jets are observed in FR-II sources.\\

We suggest that the source is a radio galaxy with asymmetric jets and lobes, probably caused by Doppler boosting and interaction with the ambient medium, and with a morphology indicating a FR-I galaxy. However, the morphological classification of this source is challenging due to the variety of characteristics observed. There are no conclusive morphological elements to clearly classify the source as a FR-I or FR-II, therefore we can not exclude the FR-II scenario.\\

The global spectral index distribution is steeper ($\alpha \sim -1.1$) than that observed in most active radio galaxies. The steep diffuse emission that is fairly bright at LOFAR frequencies has not been detected at higher frequency, which allows us to limit the spectral index to $\alpha < -1.8\pm 0.2$. We interpret this emission as lobe emission seen in projection, i.e old plasma from the two jets that were forced to bend (or old plasma left behind) by interacting with the ambient medium. The steep spectrum of the inner regions indicates that the particle energy content there is dominated by the low-energy electron population emitting below the sensitivity limit of the GMRT. In this scenario, the oldest plasma is located in the inner regions of the radio galaxy.\\

\begin{figure}
\centering
\includegraphics[width=.5\textwidth]{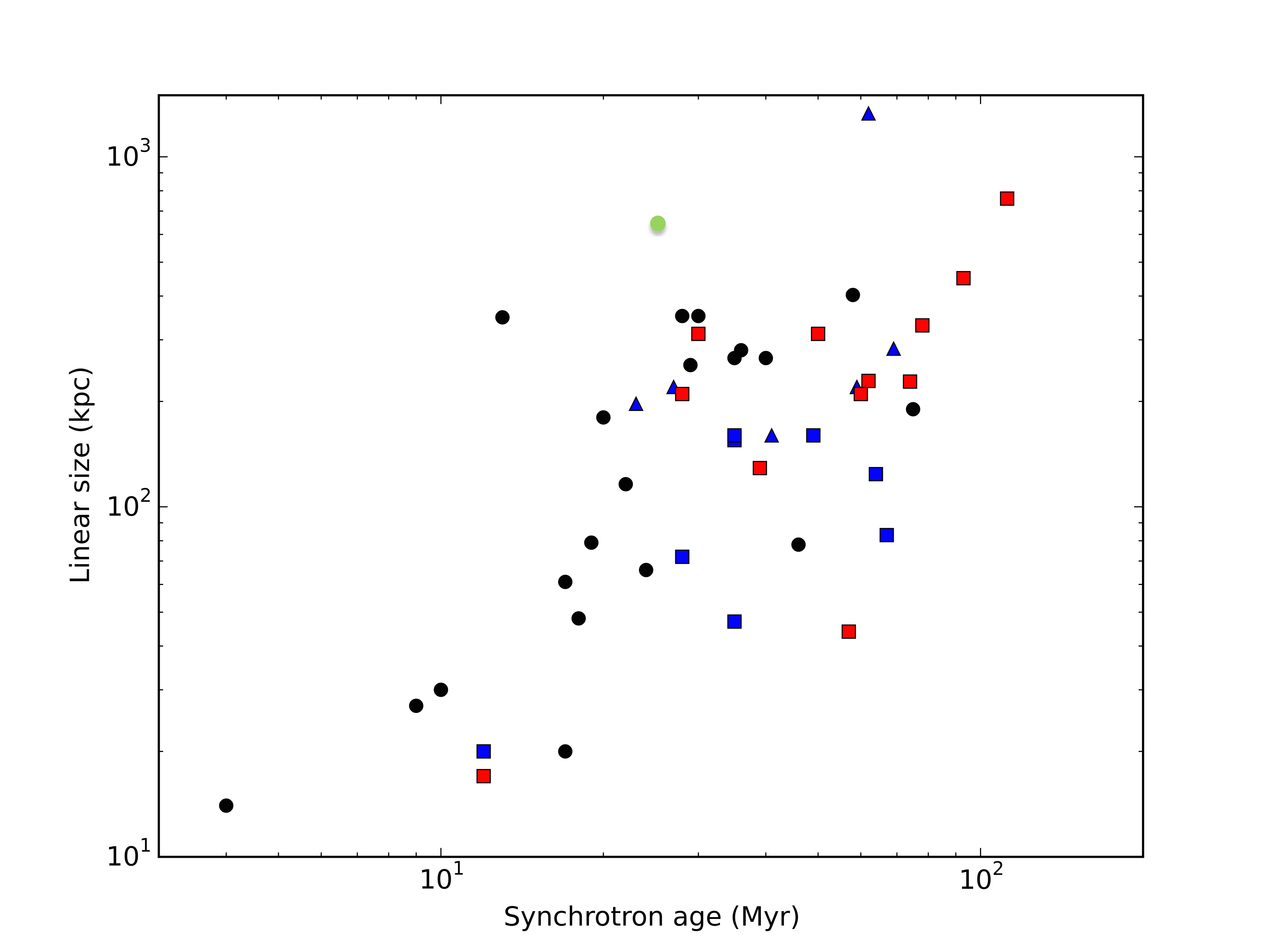}
\caption{Linear size as a function of the synchrotron age for the sample of low-luminosity radio galaxies in \citealp{Parma1999}. We added the values for MaxBCG J199, whose age being a lower limit is indicated by a green circle. Squares represent FRI sources, triangles FRII sources, and circles sources whose classification is not clear. The color represents the spectral type: red for type 1, blue for type 2, black for sources whose spectral classification is not clear.} 
\label{parma2}
\end{figure}

The spectral analysis shows that the inner regions have steeper spectral indices compared to the outer lobe regions (we refer to it as spectral type 2). This trend of spectral index steepening in the direction of the core region has already been found in both FR-I and FR-II radio galaxies (e.g. \citealp{Parma1999}). When the steepening occurs from the core outward, we refer to it as spectral type 1. We compared the linear size $LS$ and synchrotron age of MaxBCG J199 computed via the CI$_{\rm off}$ model with those of low-luminosity radio galaxies (both FR-I and FR-II) in the sample selected by \cite{Parma1999}. As shown in Fig. \ref{parma2}, MaxBCG J199 (indicated by a green circle) lays within the correlation $LS  \propto t_s^{0.97 \pm 0.17}$. We note that the linear size of MaxBCG J199 is larger than most of the galaxies in the sample, placing it in the upper region of the correlation plot.\\

The overall integrated flux density (3$\sigma$ NVSS) at 1.4 GHz is $S_{\rm tot, 1.4} = 21 \pm 3$ mJy, corresponding to a total radio power of $P_{\rm tot, 1.4} \sim 2.1 \times 10^{24}$ W Hz$^{-1}$ and the integrated flux density of the core region (3$\sigma$ FIRST) is $S_{\rm core, 1.4} = 1.3 \pm 0.2$ mJy, corresponding to a total radio power of $P_{\rm core, 1.4} \sim 1.3  \times 10^{23}$ W Hz$^{-1}$.\\
The ratio $R$ of core radio power at 1.4 GHz to total flux density at 150 MHz is defined as core prominence, and is used by a few authors as a criterion to search for remnant sources (e.g.  \citealp{Hard2016}). When $R < 10^{-4} - 5 \times 10^{-3}$, it might indicate a remnant source. 
However, this method alone is not enough to select remnant sources efficiently.
The core prominence of our source is $R = P_{\rm core, 1.4} / P_{\rm tot, 1.4}  \sim 6 \times 10^{-2}$, which does not place the source in the remnant range.\\

\begin{figure*}
\centering
\includegraphics[width=1\textwidth]{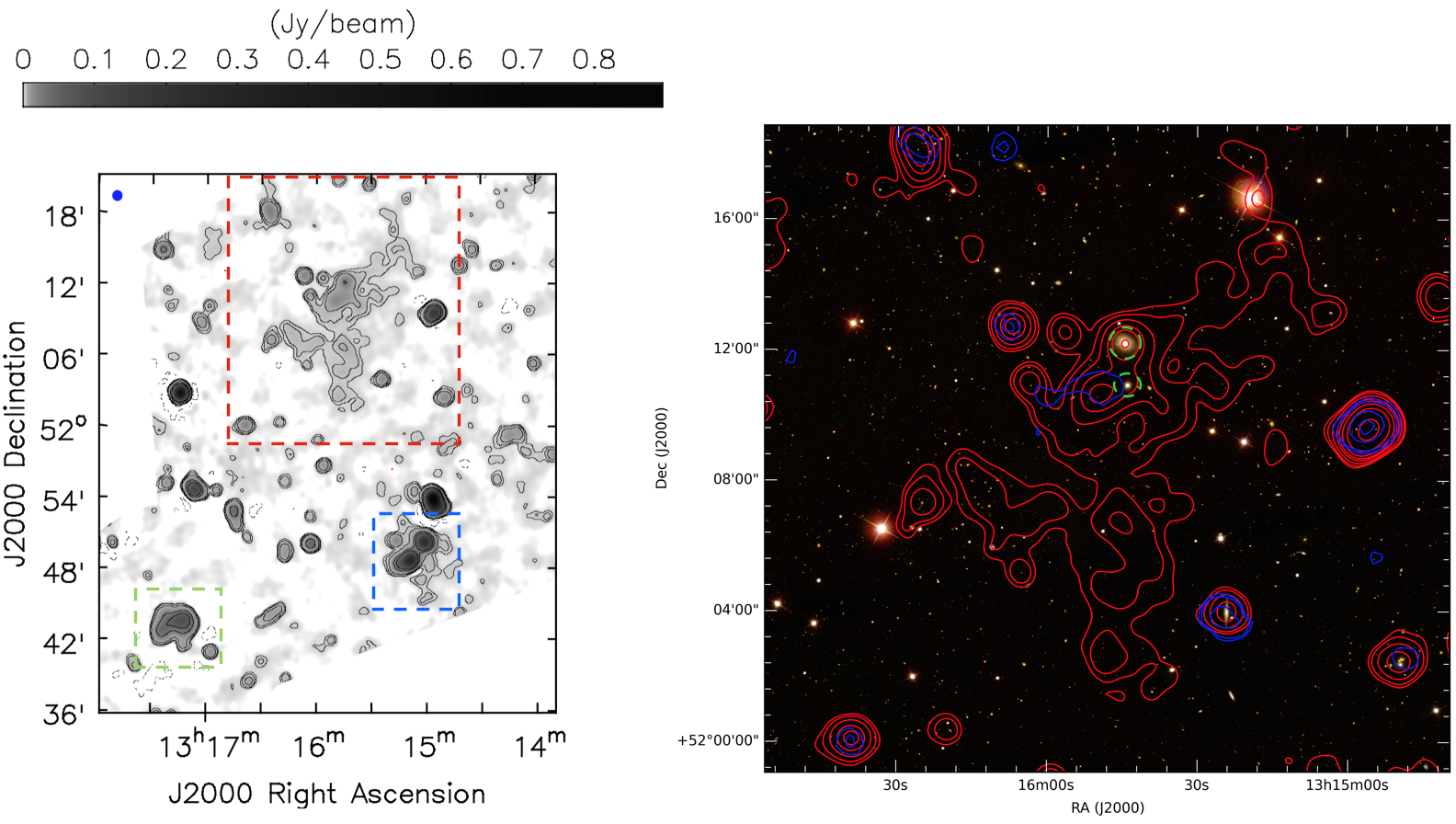}
\caption {{\bf Left panel:} Target facet image at  the central frequency 144 MHz obtained with the Briggs scheme \citep{Briggs}, robust=0.2, and 30$''$ taper. The contour levels are at $(-3, 3, 5, 10, 50, 100)\, \times \sigma$ where $\sigma$ = 450 $\mu$Jy/beam. The beam shown in blue at the top left is 50$'' \times $47$''$.  MaxBCG J199 (green square) is at the south-east, A1703 (blue square) at the south west and the new diffuse source (red square) extends to the north side of the facet.
{\bf Right panel: } SDSSg,r,i mosaic image of the diffuse source with radio contours from NVSS in blue and LOFAR in red. NVSS contour levels are at $(3, 5, 20, 40)\, \times \sigma$ where $\sigma$ = 440 $\mu$Jy/beam. LOFAR contour levels are at $(3, 5, 10, 20, 30, 50, 100, 300)\, \times \sigma$ where $\sigma$ = 450 $\mu$Jy/beam. The two galaxies at a redshift of $z = 0.0587 \pm 2 \times 10^{-4}$ that might be associated with this radio emission are indicated with green dashed circles.}
\label{new}
\end{figure*}

\begin{figure*}
\centering
\includegraphics[width=0.9\textwidth]{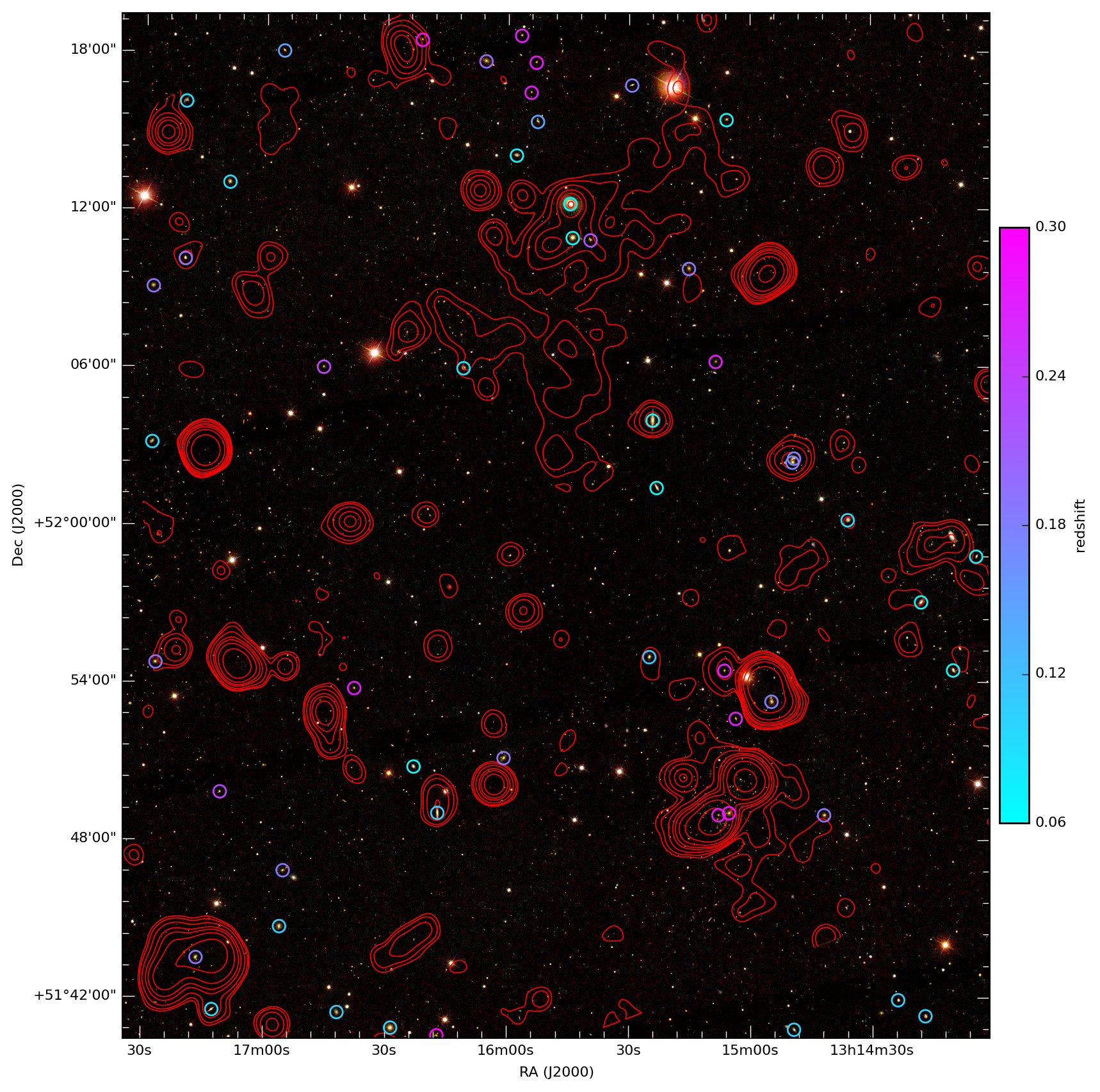}
\caption{SDSSg,r,i mosaic image of the target facet. LOFAR contour levels are at $(-3, 3, 5, 10, 50, 100)\, \times \sigma$ where $\sigma$ = 450 $\mu$Jy/beam. Galaxies with known spectroscopic redshift are marked with circles colored by their corresponding redshift.} 
\label{redshift}
\end{figure*}

MaxBCG J199 is an evolved source: material was transported up to hundreds of kpc and then travelled back toward the inner regions. The youngest plasma is located at the outer lobe edges and regions closer to the host galaxy become progressively older with the oldest material is found close to the core region. However, the compact core of radio galaxies has usually a flat spectrum, on the contrary of what happens in MaxBCG J199 where it shows a steep spectral index. Therefore, we consider two possible scenarios: 
\begin{itemize}

\item the source is still active and the core region has a spectrum steeper than standard active radio galaxies; this could be explained assuming that the steep-spectrum emission from the lobes is preventing us from resolving the core; the active core could be flat ($\alpha \sim 0$) or steep ($\alpha \sim -0.7$; e.g. \citealp{LB2014}). Assuming the lower limit $\alpha \sim -0.7$ and extrapolating from the 1.4 GHz FIRST flux density ($S_{\rm core, 1.4}$), we derived the upper limit on the core emission at 144 MHz to be 6.4 mJy. As expected, the result is lower than the measured value $S_{\rm core, 144} = 8.9 \pm 1.3$ mJy.\\

\item the source could be a dying AGN with the radio emission at the position of the host galaxy being the oldest; the jets and the core are still detectable, but the synchrotron spectrum is steepening towards the inner region that has recently stopped supplying fresh particles through the nuclear activity. Moreover, the core region that includes the base of the NW jet is likely to be beamed.

The spectral trend of the core region supports the dying-scenario, as well as the overall steep spectral value ($\alpha \sim -1.1$). \citet{Har2017} shows that there is a significant difference in the spectrum of active and remnant sources and the models used to fit the spectrum can be a good indicator of a source's current state. The model fitting of the spectrum of MaxBCG J199 gives a significantly smaller chi-square value for the $CI_{\rm off}$ model. However, the dying-scenario is not fully convincing since the core region can be seen up to 1.4 GHz, even though showing a low level emission. A more detailed study of the spectrum that would allow for instance a resolved fitting (see \citealp{Har2017}) is needed to better understand the origin of the source and its diffuse emission.

\end{itemize}

Under simple assumptions, we derived the total radiative ages of the source despite of the problems with CI models applied to radio galaxies. The age estimated through the CI$_{\rm off}$ model is comparable with the ages of known dying radio sources ($10^7 - 10^8$ Myr, \citealp{Giaci2007}, \citealp{Parma2007}, \citealp{Murgia2011}, \citealp{Brie2016}). We note that the dying radio sources presented in \citealp{Parma2007}, \citealp{Murgia2011}, and \citealp{Giaci2007} have been studied at higher radio frequencies than LOFAR and they all have linear sizes < 230 kpc, which is much smaller than the size of the radio source in MaxBCG J199. In \citet{Brie2016} the discovery of a 700-kpc remnant radio galaxy is reported. The oldest and diffuse emission can be only seen up to 1.4 GHz.\\

With the present observations, we can not exclude other scenarios. A possibility is that we might also be observing a second episode of activity. Identifying restarting/intermitting AGN is usually related to morphological features, such as multiple radio lobes, the motion of the core or hints of a jet precession. The AGN dormant phase, i.e. the period of inactivity between two episodes, could last from several Myr to tens of Myr. In our case, the source might have switched on after a rotation of the jet axis, leaving a wing in the south-west lobe as a result of the first active phase. In this scenario, the old plasma seen in projection is the aged large-scale structure with an embedded restarted radio source. Distinct episodes of AGN activity in a radio galaxy have already been observed with LOFAR, such as recurrent AGN activity in \citet{Shu2015} or an AGN relic with a restarted core in \citet{Brienza2016}. Moreover, the multiple core system might contain more than one AGN, each one of them with a different duty cycle.

\section{A suspected remnant source}
\label{sec:susp}

Two further interesting sources can be found close to MaxBCG J199: the lensing, X-ray luminous galaxy cluster A1703 (13:15:06.6, +51:49:29; z = 0.281), which will be treated in a separate paper (Savini et al., in prep), and a new radio source (13:15:44.0, +52:10:55.7), which has previously never been detected. This source is clearly visible only in the LOFAR low-resolution images that are more sensitive to diffuse emission than higher-resolution (non-tapered) images. The source is shown in Fig. \ref{new} and has an integrated flux density of $S_{144} \sim$ 600 mJy. \\
NVSS, WENSS, and other radio surveys do not show any extended diffuse emission in that region of the sky. The extent of the source is $\sim$ 1200$''$ (20$'$). We note that the largest angular scales that NVSS (VLA in D configuration, 1.4 GHz; overlay in right panel in Fig. \ref{new}) is 970$''$, therefore this source may be partially or completely resolved out in those surveys. The largest angular scale that FIRST and VLSSr (VLA in B configuration, snapshots at 1.4 GHz and 74 MHz, respectively) can observe is 60$''$ $\sim$ 1100$''$. Only the latter is comparable to the extent of the source. However, no diffuse emission is present, probably due to the very low sensitivity of these snapshots surveys.
The GMRT minimum baseline is 150$\lambda$, which corresponds to a maximum detectable scale of 1375$''$ (23$'$), which is comparable to the extent of the source (with the caveat that the inner {\it uv} coverage could be not sufficient). However, we do not detect any emission in our GMRT observation centered on MaxBCG J199. \\
Hence, no upper limits on the spectrum can be computed. Observations, such as a LOFAR LBA pointing will help to determine the nature and morphology of this peculiar emission.\\

We searched in the NASA/IPAC Extragalactic Database (NED)\footnote{https://ned.ipac.caltech.edu} for possible counterparts to this diffuse emission. In Fig. \ref{redshift}, the position of close-by galaxy clusters and galaxies with known redshift is shown.  We found no massive galaxy clusters in the region of the source within an angular radius of 0.5$^{\circ}$. The two galaxies SDSS J131544.56+521213.2 and SDSS J131543.99+521055.7 at the spectroscopic redshift of $z = 0.0587 \pm 2 \times 10^{-4}$ (\citealp{SDSS2007}) might be associated with the radio emission. The emission detected by LOFAR above 10$\sigma$ seems to be centered on these two galaxies, which are indicated with dashed circles in the right panel in Fig. \ref{new}. Therefore, we speculate that the faint diffuse radio emission in its entire extent is connected to the two galaxies with a projected separation of $\sim$ 79$''$ that corresponds to $\sim$ 90 kpc. They are classified as an AGN pair by \cite{Liu2011}, where interacting AGN pairs with separations from kpc to tens of kpc  are optically selected from SDSS (Data Release 7; \citealp{SDSS2009}). This implies that the supermassive black holes (SMBH) in their nuclei are active during the same stage of a galaxy-galaxy merger, and accretion onto the SMBH and host-galaxy star formation is enhanced by the galaxy tidal interactions \citep{Liu2012}. At the spectroscopic redshift of this AGN pair, the extension of the source would be $\sim$ 1 Mpc. The northern member of the AGN pair is located at the peak of the radio emission, while the second member is offset from the second peak of the radio emission.
One possible scenario is that the optical galaxies are actually dying radio galaxies which interacted as an AGN pair in the past, and whose radio emission might be old and steep. The lobes are fading away, and the emission can be interpreted as a radio remnant source. \\
We note that the shape of this source is comparable with the dying radio galaxy WNB 1851+5707a seen at 1.4 GHz, although the latter extends on a much smaller scale (Fig. 6 in \citealp{Murgia2011}).

\section{Summary}
\label{sec:conc}
We present the discovery of extended radio emission at LOFAR frequencies at the coordinates of the galaxy group MaxBCG J199 (RA = 13:17:16.4, DEC = +51:43:30.0, J2000). SDSS photometric data reveal a total of 13 galaxies within this group and a multiple core system composed of three galaxies.
We performed the reduction of the LOFAR data using the facet calibration method to reach a rms noise of 135 $\mu$Jy/beam and resolution of 10.6$'' \times$ 6.0$''$ at HBA frequencies (120 - 168 MHz). LOFAR observations show that the radio diffuse emission is connected to a central radio galaxy whose powerful radio jets and lobes extend on angular scales of 3.4$'$, corresponding to a linear size of 650 kpc at the spectroscopic redshift of the source.  The core of the radio galaxy is coincident with the brightest galaxy of the group at $z = 0.18793 \pm 5 \times 10^{-5}$.\\
We obtained a GMRT follow-up observation at 607 MHz to study the spectral properties of the sources. GMRT images reach a rms noise of 60 $\mu$Jy/beam at 6.0$'' \times$ 4.8$''$ resolution. LOFAR images show a greater extent than GMRT emission: only the brightest and more compact emission is visible in the GMRT image, which can be used to obtain a low-frequency spectral index map. The spectral index values range from -1.3 to -1.1 in the core region and inner edges, and from -0.7 to -0.5 at the outer edges, therefore it steepens going towards the inner regions. The global spectral index value of the source calculated using the integrated flux densities above 3$\sigma$ from GMRT and LOFAR is relatively steep, around -1.1. The diffuse emission detected at 120 - 168 MHz that can not be seen at 591 - 623 MHz must be steeper, and we place an upper limit of $-1.8 \pm 0.2$. We interpret this emission as old lobe emission seen in projection. 
The extension of the source suggests a strong nuclear activity of the central engine. The low-frequency spectral index map obtained between LOFAR and GMRT images reveals a steepening of the spectrum from the lobe outer edge inward and a steep core, which is in disagreement with the usual spectrum of active nuclei. Moreover, the spectral index map indicates activity only at the edge of the lobes. Therefore, we considered two possible interpretations: the source is active, but we observe a mix of core and steep spectrum emission that causes the spectral index of the core to appear steeper than it really is; or the source is dying, i.e. the AGN has recently entered a phase of quiescence, where the nucleus stopped supplying fresh electrons to the lobes.\\
We conclude that the radio source found in MaxBCG J199 is an evolved radio-loud AGN surrounded by diffuse emission that can be best studied with LOFAR and is likely related to old plasma left behind by the jets forming the lobes or a continuation of the lobes that experienced a backflow at the edges. The two jets are interacting with the intragroup medium that shows asymmetrical features in the lobe regions.\\
These observations probe the great potential of LOFAR to detect old plasma, and demonstrate that low-energy electrons are present in the intragroup medium, and could furnish a seed population for particle re-acceleration mechanisms.
The source that we have presented in this paper is an example of steep-spectrum radio source that low-frequency surveys, such as LoTSS, can discover.

\section*{Acknowledgments}

LOFAR, the Low Frequency Array designed and constructed by ASTRON, has facilities owned by various parties (each with their own funding sources), and that are collectively operated by the International LOFAR Telescope (ILT) foundation under a joint scientific policy.\\ 
We would like to thank the staff of the GMRT that made the observation possible. GMRT is run by the National Centre for Radio Tata Institute of Fundamental Research.\\ 
The research leading to these results has received funding from the European Research Council under the European Union's Seventh Framework Programme (FP/2007-2013) / ERC Advanced Grant RADIOLIFE-320745. M. J. Hardcastle acknowledges support from the UK Science and Technology Facilities Council [ST/M001008/1]. P. N. Best is grateful for support from the UK STFC via grant ST/M001229/1. A. O. Clarke gratefully acknowledge support from the European Research Council under grant ERC-2012-StG-307215 LODESTONE.\\
This research made use of the NASA/IPAC Extragalactic Database (NED), which is operated by the Jet Propulsion Laboratory, California Institute of Technology, under contract with the National Aeronautics and Space Administration.\\
This research made use of APLpy, an open-source plotting package for Python hosted at http://aplpy.github.com.\\
F. Savini thanks Steven N. Shore (University of Pisa) for his helpful comments.



\bibliographystyle{mnras}
\bibliography{paperLOFAR-MNRAS-new}







\bsp	

\label{lastpage}
\end{document}